# 'Bosons' and 'fermions' in social and economic systems

## Sergey A. Rashkovskiy

*Ishlinsky Institute for Problems in Mechanics of the Russian Academy of Sciences, Vernadskogo Ave., 101/1,*

*Moscow, 119526, Russia*

*Tomsk State University, 36 Lenina Avenue, Tomsk, 634050, Russia*

*E-mail: rash@ipmnet.ru, Tel. +7 906 0318854*

**Abstract** We analyze social and economic systems with a hierarchical structure and show that for such systems, it is possible to construct thermostatistics, based on the intermediate Gentile statistics. We show that in social and economic hierarchical systems there are elements that obey the Fermi-Dirac statistics and can be called fermions, as well as elements that are approximately subject to Bose-Einstein statistics and can be called bosons. We derive the first and second laws of thermodynamics for the considered economic system and show that such concepts as temperature, pressure and financial potential (which is an analogue of the chemical potential in thermodynamics) that characterize the state of the economic system as a whole, can be introduced for economic systems.

**1. Introduction**

Recently, the possibility of the application of physical techniques to areas outside of physics is actively discussed in the literature. We would like to highlight the two of the most intensively developing areas: (i) application of techniques and approaches of quantum mechanics to the description of various social systems [1,2] and (ii) application of techniques and approaches of statistical physics and thermodynamics to the description of economic systems [3-9].

The search in these areas led to finding a number of interesting similarities between physical systems on the one hand and social and economic systems on the other.

These similarities are not accidental; they have deep roots. Indeed, in the simplest case, the social (economic) system, at least formally, can be compared with gas: in both cases the system consists of a set of interacting elements that are the minimal structural units of the system. In the gas, the elements are atoms and molecules; in the social (economic) system, the elements are individuals or their associations (family, firm, company, party, settlement, state, etc.). In both cases, the elements can be either simple (indivisible in the process under consideration) or complex, having an internal structure and capable of decaying into simpler ones, or, conversely, uniting into more complex ones.

In both cases, randomness plays an important role in the behavior of the system and leads to the need to use statistical methods to describe it. The main cause of randomness in all these systems is the absolute instability of the "phase trajectories" of the individual elements, which leads to the emergence of dynamic chaos, even in a deterministic system. Here we would like to draw attention to another similarity between social (economic) systems and quantum systems in physics: the randomness in the behavior of the elements of these systems is associated not only with dynamic chaos and random external impacts, but also with an internal randomness inherent in the elements themselves. So, the random behavior of physical objects (for example, electrons and atoms) is related with the very quantum nature of these objects. Similarly, the behavior of an individual in a social (economic) system is also largely random and is associated with his internal (physical and mental) state.

In both cases, one of the most important driving forces of system development is competition for the possession of limited resources. If for social and economic systems this statement seems obvious (a limited resource is the wealth available in the system), then for physical systems this statement requires an explanation. Consider a gas in a closed vessel which has a constant energy. In the collision (interaction) of atoms (molecules) in the gas, an energy exchange occurs between them: the atom (molecule) either gains additional energy, or loses some of its energy. As a result, there is a constant redistribution of energy between atoms (molecules). This can be seen as a competition between atoms (molecules) for possessing a limited resource, which is the energy of the whole gas.

All this leads to the fact that the specific mechanisms of interaction of elements in such systems (atoms and molecules in physical systems or agents of social and economic systems) are not determinative, and certain common collective properties of elements come to the fore. As a result, all such systems, regardless of their nature, have common properties and obey the same principles.

This is what allows us to hope that many of the techniques and approaches developed for physical systems can also be applied to social and economic systems.

The aim of this paper is to analyze yet another informal similarity between physical and economic (social) systems: the existence of elements in social and economic systems possessing the properties of bosons and fermions.

**2. Energy and states of economic systems**

Money in the economic system plays the same role as internal energy in a physical thermodynamic system [3-9]. The energy of an ideal gas is composed of the kinetic energy of all the atoms and molecules that make up this gas. Similarly, the amount of money available in the economic system is equal to the sum of all the money that the elements of this system have. For money, as for energy, there is a conservation law. Indeed, let us consider two elements of the economic system. These elements can interact with each other. By interaction we mean any process in which the elements exchange money.

For example, one element sells some good or service, while another buys it, or simply, the elements transfer money to each other (lend, give, etc.). Let the amount of money that the elements had before the interaction is, respectively, $\varepsilon_1$ and $\varepsilon_2$. As a result of the interaction of these elements, the amount of money for each of them will change and become, respectively, $\varepsilon'_1$ and $\varepsilon'_2$. In this case, obviously,

$$\varepsilon_1 + \varepsilon_2 = \varepsilon'_1 + \varepsilon'_2 + \delta\varepsilon \tag{1}$$

where $\delta\varepsilon$ is the amount of money that was passed to the third elements (third parties) as a result of this interaction. The parameter $\delta\varepsilon$ describes possible losses (taxes, physical loss of money, etc.) or the gain during this financial transaction.

Similarly, we can write the law of conservation of "energy" for two interacting economic systems

$$E_1 + E_2 = E'_1 + E'_2 + \delta E \tag{2}$$

where $E_1, E_2$ are the amount of money ('energy') in systems 1 and 2 before the interaction; $E'_1, E'_2$ are the amount of money in systems 1 and 2 after interaction; $\delta E$ is the amount of money ('energy') lost or gained by systems 1 and 2 as a result of interaction with the third system (taxes, investments, etc.).

Expressions (1) and (2) are similar to the law of conservation of energy for physical (e.g. thermodynamic) systems.

In other words, money in economic systems, similarly to energy in physical systems, does not disappear anywhere and does not appear out of nowhere; they can only go from one system (element) to another or from one form to another (from one currency to another, from a cash form to a cashless form, from a cash form to a good, etc.). In any case, we can always compose a financial balance that will take into account all the channels of money ('energy') transfer, and which will converge absolutely. The role of money, as the energy of the economic system, was discussed in one form or another in [3-9].

Hereinafter, the term 'energy' as a synonym for the term 'amount of money' as applied to economic systems will be used without quotes.

By analogy with the physical thermodynamic system, we will assume that the probability of any state of the economic system is determined by the energy of this state (i.e., by the amount of money that the system has in this state): $p = p(E)$.

Consider an economic system consisting of two subsystems 1 and 2. The energy of this system will be an additive quantity:

$$E = E_1 + E_2 \tag{3}$$

where $E_1$ and $E_2$ are the amount of money in subsystems 1 and 2.

Probability of the state of the whole system is $p = p(E)$. Probabilities of states of subsystems 1 and 2 are $p_1 = p(E_1)$ and $p_2 = p(E_2)$.

Assuming that the subsystems are statistically independent, one obtains $p = p_1 p_2$ or

$$p(E_1 + E_2) = p(E_1) p(E_2)$$

From this, as usual, one obtains the Gibbs distribution

$$p(E) = Z^{-1} \exp(-\beta E) \tag{4}$$

where $\beta$ is the some characteristic of the system, which is the same both for the system as a whole and for each of its subsystems; $Z$ is the partition function.

Taking into account that $\sum_E p(E) = 1$, one obtains

$$Z = \sum_E \exp(-\beta E) \tag{5}$$

where the summation is over all possible states having different energies. If we take into account degenerate states, then

$$p(E) = Z^{-1} g(E) \exp(-\beta E) \tag{6}$$

where

$$Z = \sum_E g(E) \exp(-\beta E) \tag{7}$$

where $g(E)$ is the degree of degeneracy of the states with energy $E$ (that is, the number of different states of a system having the same energy).

In what follows we confine ourselves to the case $g(E) = 1$.

If the economic system has a variable number of elements $r$, then its state is characterized by two parameters: $E$ and $r$. Therefore, the probability of the state of the economic system is $p = p(E, r)$.

If the system consists of two statistically independent subsystems, then $E = E_1 + E_2$, $r = r_1 + r_2$ and $p = p_1 p_2$ or $p(E_1 + E_2, r_1 + r_2) = p(E_1, r_1) p(E_2, r_2)$

From this, as usual, one obtains the Gibbs distribution

$$p(E,r) = Z^{-1} \exp(\alpha r - \beta E) \quad (8)$$

where $\alpha$ is the parameter which characterizes the economic system and which is the same both for the system as a whole and for each of its subsystems.

Taking into account that $\sum_{E,r} p(E,r) = 1$, one obtains

$$Z = \sum_{E,r} \exp(\alpha r - \beta E) \quad (9)$$

where the summation is over all possible states of the system having different energies and different number of elements.

Let us consider the case when the energy of an economic system is uniquely determined by the number of its elements and is additive in the number of elements, i.e.

$$E = \varepsilon r \quad (10)$$

This means that all elements in such an economic system have the same amount of money (energy) $\varepsilon$.

Then

$$Z = \sum_r \exp((\alpha - \beta \varepsilon) r) \quad (11)$$

where the summation is over all possible values of $r$.

The mean number of elements in such a system (the mean population of economic system)

$$\langle r \rangle = \sum_r r p(r) \quad (12)$$

In particular for the case (10), one obtains $\langle r \rangle = Z^{-1} \sum_r r \exp(\lambda r)$, where $\lambda = \alpha - \beta \varepsilon$.

In this case, the mean population of the economic system, as usual, is easily calculated using the partition function

$$\langle r \rangle = \frac{d \ln Z}{d \lambda} \quad (13)$$

## 3. Fermions in economic systems

In quantum mechanics, there are two types of particles: fermions and bosons. Fermions obey the Pauli exclusion principle: in one state there cannot be more than one particle, while for bosons there are no restrictions on the number of particles in one state.

The question arises: are there elements in economic or social systems that obey the principle similar to the Pauli exclusion principle?

We can answer this question in the affirmative, since it is very easy to find the relevant examples.

Indeed, the position of the head of various levels (the president of the country, the prime minister, the president of the company, the director of the company, etc.) can only be occupied by one person. From this point of view, all applicants for these positions will be fermions in the sense that two or more of them will never be able to occupy the same position at the same time.

Similarly, we can consider a goods that can have only one owner. Suppose, for example, that the owner of an apartment can be only one person (we do not consider here the case when several individuals can own the same apartment in different proportions). In this case, the apartment offered for sale may be either vacant or occupied (bought) by only one person. From this point of view, all potential buyers of apartments are fermions in the sense that two or more of them can never be the owners of the same apartment at the same time.

Thus, we can formulate a general principle - the 'Pauli exclusion principle' for social and economic systems: in social and economic systems there are states that can either be vacant, or can be occupied by only one element of the system. All applicants for these states can be called fermions.

Consider in more detail the idealized housing market, in which the similar apartments with the similar characteristics (the same quality, the same conditions, the same location, etc.) are sold, the owners of each of which can be only one person. Let the cost of the apartment is equal to $\varepsilon$ and the same for all the apartments under consideration.

Each of these apartments can be considered as a system that has two permissible states: vacant and occupied (bought). In addition, there are elements-fermions (potential buyers of apartments) that can occupy these systems.

The state of such a system (apartment) is characterized by energy $E = \varepsilon r$, where $r$ is the number of apartment owners: $r = 0$ (the apartment is vacant) and $r = 1$ (the apartment is bought, occupied).

The probability of the state of such a system is described by the Gibbs distribution (8), (10). Then according to (11),

$$Z = 1 + \exp(\alpha - \beta\varepsilon) \tag{14}$$

Obviously, the mean population $\langle r \rangle$ of the system under consideration has a simple and visual meaning: it is equal to the share of occupied (bought) apartments in the housing market.

Using (13), for the mean population of the system, we obtain

$$\langle r \rangle = \frac{1}{\exp(\beta\varepsilon - \alpha) + 1} \tag{15}$$

Thus. the mean population of apartments is described by the Fermi-Dirac distribution (Fig. 1).

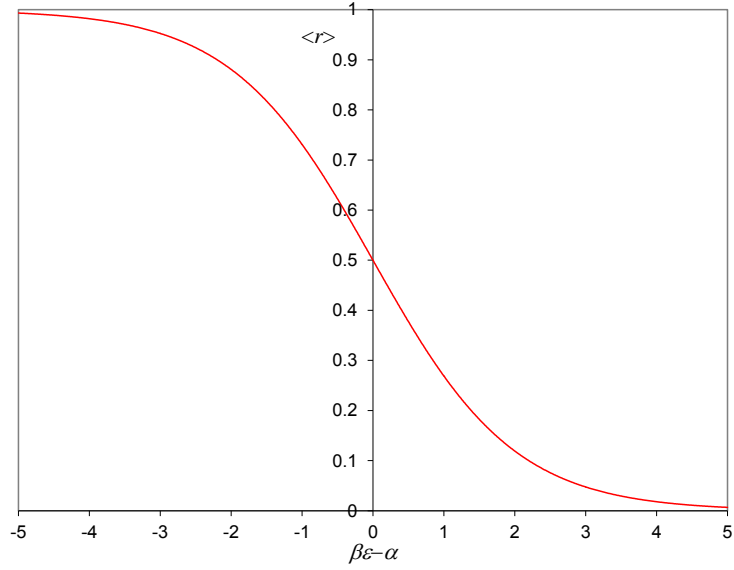

Fig. 1. Share of occupied (bought) apartments in the housing market (mean population of the system), depending on their cost $\varepsilon$.

As we would expect, with the increase in the cost of apartments, the share of apartments sold (the mean population of the system) decreases. On the contrary, at a low cost of apartments ($\varepsilon \ll \alpha/\beta$) the share of apartments sold (the mean population of the system) tends to unity.

Actually, even absolutely identical apartments can have different costs. We introduce the distribution of identical apartments at a cost: $\varphi(\varepsilon)$, where $\varphi(\varepsilon)d\varepsilon$ is the relative number of identical apartments, the cost of which is in the range $[\varepsilon, \varepsilon + d\varepsilon]$. Then the mean population of apartments of the type under consideration in the housing market

$$\langle r \rangle = \int_0^\infty \frac{\varphi(\varepsilon)d\varepsilon}{\exp(\beta\varepsilon - \alpha) + 1} \tag{16}$$

Let us consider another example. Let there are many identical firms (companies), each of which is headed by a director. The position of the director can be occupied by only one person, therefore the company can be considered as a system in which there can be no more than one element. In addition, we assume that there are many elements-fermions (people) claiming this position and having the same characteristics (education, work experience, skills, age, etc.).

This position is characterized by the level of salary $\varepsilon$ which determines the energy of the system. However, unlike the previous example with apartments, salary should be considered as potential energy of the system. Indeed, the higher the salary of the director, the more applicants for this position, the higher the probability that this state will be occupied. This is similar to a potential well in physical systems: the deeper the potential well, the less probability that the particle will leave it, the higher the probability that this state will be occupied. Continuing the analogy, we can say that the salary of the director is similar to the depth of the potential well. Thus, in this case we should take

$$E = -\varepsilon r \qquad (17)$$

where $r$ is the number of elements (people) that occupy this position.

The system under consideration can be in two states: $r=0$ (the position of director is vacant) and $r=1$ (the position of director is occupied). The probabilities of these states are described by the Gibbs distribution (8), which we rewrite in the form

$$p(r) = Z^{-1} \exp\left((\beta\varepsilon + \alpha)r\right) \qquad (18)$$

Then

$$Z = 1 + \exp(\beta\varepsilon + \alpha) \qquad (19)$$

Obviously, the mean population $\langle r \rangle$ of the system under consideration also has a simple and visual meaning: it is equal to the share of the occupied positions of the director in the labor market.

Using (13), for the mean population of the position of director, we obtain

$$\langle r \rangle = \frac{1}{1 + \exp(\alpha - \beta\varepsilon)} \qquad (20)$$

which formally is also the Fermi-Dirac distribution (Fig. 2).

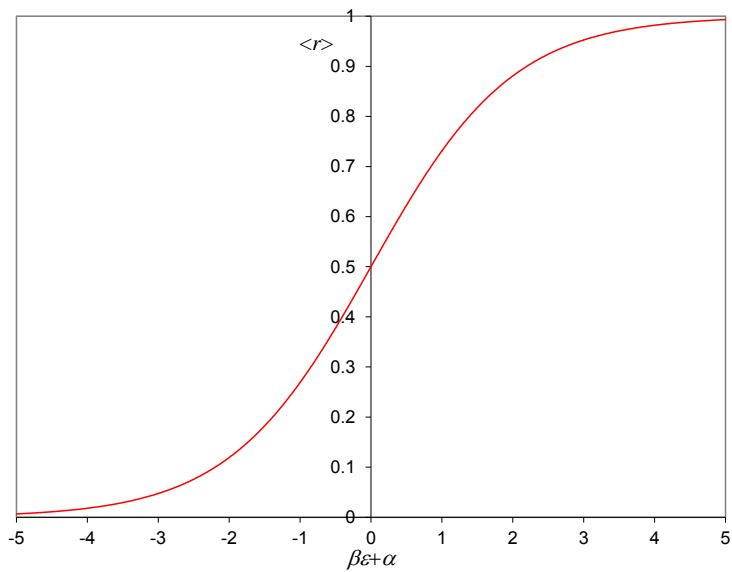

Fig. 2. Share of the occupied positions of the director in the labor market (mean population of the system) depending on the offered salary $\varepsilon$.

We note that the parameter $1-\langle r \rangle$ is equal to the share of vacant positions of the director and is described by the conventional Fermi-Dirac distribution.

As we would expect, with the growth of the offered salary, the share of occupied positions of directors of companies in the labor market (the average population of the system) increases. On the contrary, with a very low salary offered ($\varepsilon \ll -\alpha/\beta$), the share of the occupied positions of the director in the labor market (the mean population of the system) tends to zero.

## 4. Statistics of hierarchical systems

Most social and economic systems have a hierarchical structure. They consist of several levels $i=1,...,L$. At each level, there are a certain number of positions that can be occupied by the elements of the system. One position can be occupied by only one element or it can be vacant. The number of available positions on the level $i$ is $d_i$. Thus, from zero to $d_i$ elements can be simultaneously at the $i$th level of the hierarchical system. For simplicity, we assume that all positions at a given level are the same (indistinguishable). The system is hierarchical if $d_i < d_{i+1}$.

As an example, we consider the structure of the company. At the head of the company, there is a director who has a salary $\varepsilon_1$. This position corresponds to the level $i=1$ for which $d_1=1$. The director has $d_2$ of his deputies (conditionally, heads of departments), whom we consider are absolutely equal and having the same salary $\varepsilon_2$. These positions correspond to the level $i=2$. Each head of the department has his deputies (conditionally, heads of divisions). These positions correspond to the level $i=3$. We assume that all heads of divisions (regardless of which department head they obey) are absolutely equal and have the same salary $\varepsilon_3$. The total number of heads of divisions in the company is equal to $d_3$. And so on. At the very last level $i=L$ of the hierarchical system there are ordinary employees of the company, the total number of which is equal to $d_L$. All employees at the level $i=L$ are also considered to be absolutely equal (regardless of the structural unit of the company in which they work) and they all have the same salary $\varepsilon_L$.

In hierarchical systems, always $\varepsilon_i > \varepsilon_{i+1}$, because it is this distribution of salary that is the driving force behind the development of social and economic systems; this condition leads to a differentiation of the social system and to the emergence of its hierarchical structure.

Depending on what indicator is considered as a level criterion, the hierarchical system can be represented either in the form of a pyramid (if the criterion is the salary and the degree of subordination of the elements), top of which (director) corresponds to upper level, while the ordinary employees are at the bottom level, or in the form of an inverted pyramid (if the criterion is, for example, the level number), which top (director) corresponds to the bottom level, while the ordinary employees are at the upper level $i = L$. Although the first representation (in the form of a pyramid) is used traditionally, hereinafter we will use the second representation (inverted pyramid). It is in this representation that we can compare the hierarchical system with well-known physical systems.

Indeed, as indicated in the previous section, salary is similar to the depth of the potential well, while the energy of the level is determined by the relation (17). In this case, the levels with a larger $i$ number (with a lower salary) have a greater potential energy.

Elements of a hierarchical system can move from one level to another, but only if there is a vacant position at a new level. Obviously, the most probable (preferred) are transitions from a higher energy level $i$ to a lower level $j < i$, where $\varepsilon_j > \varepsilon_i$. Such transitions can be called spontaneous transitions; they can occur if a vacancy has appeared at a lower energy level $j$. In addition to spontaneous transitions, the forced transitions are also possible in the hierarchical system, when elements are forced (against their will, due to external impact) to move to a higher energy level $k > i$ such that $\varepsilon_k < \varepsilon_i$.

Thus, in the case of a spontaneous transition, the element goes to a level with a lower energy (higher salary), while in the forced transition to a level with a higher energy (lower salary). Obviously, the analogue of a hierarchical system in physics is, for example, potential wells or a mechanistic model of an atom, the elements of which are electrons. An electron can spontaneously move to a lower energy level if it is vacant, and under an external impact it can be forced to move to a higher energy level, leaving the energetically more profitable state vacant.

As in the atom (potential well), the filling the hierarchical system occurs starting from the lower energy level (for example, in the company, the levels with the highest salary are filled first and only then the levels with the lower salary). For this reason, in the hierarchical system, the lowest energy levels (the levels with the highest salary) are always the most filled.

On the contrary, the most mobile in the hierarchical system are the elements of the upper energy levels (levels with the lowest salary): vacancies appear there more often, because these elements easily leave their level, either by moving to a lower energy level (a level with a higher salary) in the system, or even leaving the system (company), moving into the 'environment'.

As shown in the previous section, the director of the company is a fermion, and the corresponding level $i=1$ (the lowest energy level) is described by the Fermi-Dirac statistics (20).

Consider the statistics of other levels of the hierarchical system.

Take a level of the hierarchical system in which there are $d$ positions with the same salary $\varepsilon$. We consider this level as a system in which there can be no more than $d$ elements simultaneously. If the number of elements at this level is $r$ ($r \leq d$), then the energy of this system is determined by the relation (17), and according to (8) the probability of such state of the system is

$$p(E,r) = Z^{-1} \exp\big((\beta\varepsilon + \alpha)r\big) \tag{21}$$

where

$$Z = \sum_{r=0}^{d} \exp\big((\beta\varepsilon + \alpha)r\big) \tag{22}$$

After summation, we obtain

$$Z = \frac{1 - \exp\big((\beta\varepsilon + \alpha)(d+1)\big)}{1 - \exp\big((\beta\varepsilon + \alpha)\big)} \tag{23}$$

Using (13) and (23), for the mean population of the level under consideration of the hierarchical system, we obtain

$$\langle r \rangle = f_G(\varepsilon) \tag{24}$$

where

$$f_G(\varepsilon) = \frac{1}{\exp\big(-(\beta\varepsilon + \alpha)\big) - 1} - \frac{d+1}{\exp\big(-(\beta\varepsilon + \alpha)(d+1)\big) - 1} \tag{25}$$

Thus, the level of the hierarchical system characterized by the parameters $d$ and $\varepsilon$ is described by the intermediate Gentile statistics [10,11].

At $d=1$ (director) the distribution (25) goes over into the Fermi-Dirac distribution (20), while at $d=\infty$ it goes over into the Bose-Einstein distribution.

Obviously, the mean population of the level of a hierarchical system has a simple and visual meaning: it is equal to the mean number of occupied positions at a given level in similar hierarchical systems. As applied to companies, distribution (25) characterizes the labor market for specialists of a given level, while the value $V(d - \langle r \rangle)$ determines the number of vacancies for specialists of a given level in the labor market, depending on the offered salary $\varepsilon$, where $V$ is the number of the similar companies on the market.

The mean population of the level of the hierarchical system (24), as a function of the parameter $(\beta\varepsilon+\alpha)$ for different values of the parameter $d$, is shown in Fig. 3.

It is also interesting to consider the mean relative population $\langle r \rangle/d$ of the level of the hierarchical system that characterizes the mean share of occupied positions at the level under consideration in all identical hierarchical systems. The dependence of the function $\langle r \rangle/d$ on the parameter $(\beta\varepsilon+\alpha)$ for different values of the parameter $d$ is shown in Fig. 4.

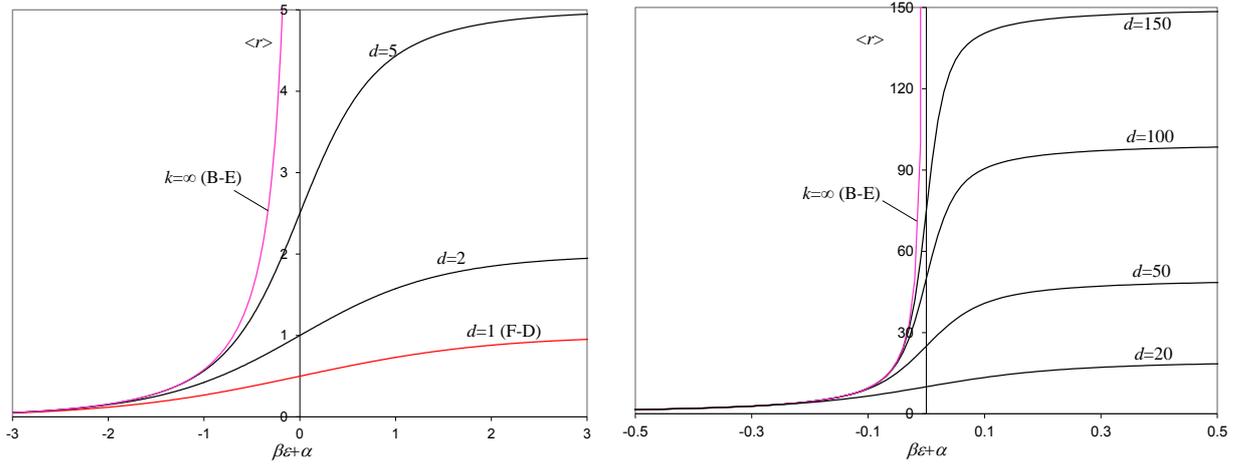

Fig. 3. Mean population of the level of the hierarchical system (24), (25) as a function of the parameter $(\beta\varepsilon+\alpha)$ for different values of the parameter $d$.

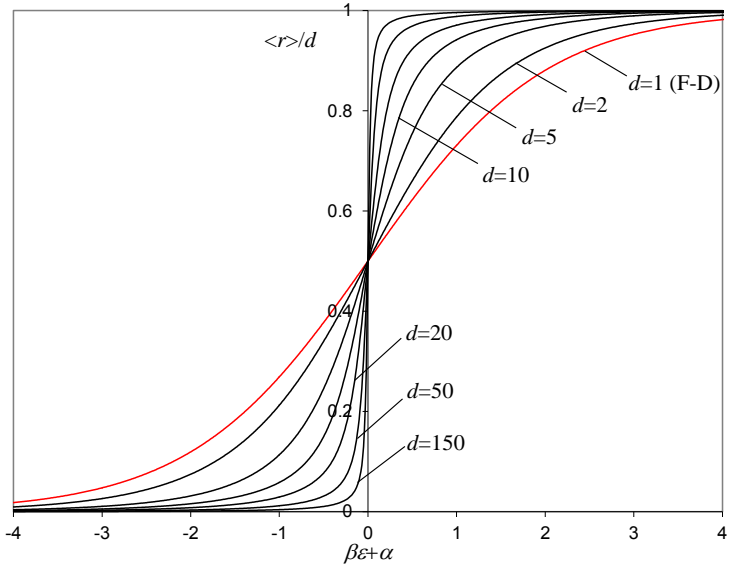

Fig. 4. Dependence of mean relative population $\langle r \rangle/d$ of the level of the hierarchical system on the parameter $(\beta\varepsilon+\alpha)$ for different values of the parameter $d$.

Thus, the levels of a hierarchical system with a large value of the parameter $d \gg 1$ are approximately described by the Bose-Einstein statistics, and the elements occupying these levels of the hierarchical system can be approximately considered as bosons. This approximation is more accurate the larger the value of the parameter $d$. In particular, ordinary employees of a large company can be considered as bosons, and the number of occupied corresponding positions in the labor market is approximately described by the Bose-Einstein distribution.

## 5. Thermostatistics of hierarchical systems

Let us consider the thermostatics of an arbitrary level of a hierarchical system based on the example of the labor market discussed in the previous section.
Hereinafter we shall follow the papers [10, 11].
Consider an ensemble of identical hierarchical systems (companies), the total number of which $V \gg 1$. We will consider only one (but the same for all companies) level, characterized by the number of available positions $d$. Generally speaking, the salary at this level for different companies may be different, but the same for all positions of this level within the same company. Let $v_r^s$ is the number of companies in which there are exactly $r$ elements ($0 \leq r \leq d$) at a given level, and the salary of each element of this level is equal to $\varepsilon_s$.
Then

$$V_s = \sum_{r=0}^{d} v_r^s \tag{26}$$

is the number of companies with any number of elements at a given level, in which the salary of each element at the level under consideration is equal to $\varepsilon_s$.
Obviously,

$$V = \sum_s V_s = \sum_s \sum_{r=0}^{d} v_r^s \tag{27}$$

The number of elements in all companies with the same salary $\varepsilon_s$ at this level is equal to

$$N_s = \sum_{r=0}^{d} r v_r^s \tag{28}$$

Then the total number of elements in all companies at this level, regardless of the salary is

$$N = \sum_s N_s = \sum_s \sum_{r=0}^{d} r v_r^s \tag{29}$$

while the energy of all companies at this level is

$$E = -\sum_s \varepsilon_s N_s = -\sum_s \sum_{r=0}^{d} \varepsilon_s r v_r^s \tag{30}$$

Conditionally we unite identical levels of all companies into one system. In other words, we are considering a system in which there are different states. These states are the same levels of different companies (hierarchical systems). Accordingly, these states can be either vacant or occupied by elements; the number of elements in each state cannot exceed $d$. Then $N$ is the total number of elements in this system, while $E$ is the energy of the system. The total number of companies $V$ we will call the volume of the system under consideration.

The number of possible ways of accommodation of elements in this system

$$W = \prod_s \frac{V_s!}{v_0^s! v_1^s! ... v_d^s!} \tag{31}$$

Having determined the Boltzmann entropy of the system under consideration, as $S = \ln W$, and using the Stirling approximation, we obtain

$$S = \sum_s V_s \ln V_s - \sum_s \sum_{r=0}^{d} v_r^s \ln v_r^s \tag{32}$$

The most probable state of the system corresponds to the maximum of entropy (32) for given values $V$, $N$ and $E$. Using the method of undetermined Lagrange multipliers, as well as the definitions (26), (27), (29), (30) and (32), we obtain

$$\sum_s \sum_{r=0}^{d} \left( \ln V_s - \ln v_r^s + \gamma + \alpha r + \beta \varepsilon_s r \right) \delta v_r^s = 0 \tag{33}$$

Where $\alpha$, $\beta$ and $\gamma$ are the undetermined Lagrange multipliers.

Hence we obtain

$$v_r^s = A_s \exp\left( (\beta \varepsilon_s + \alpha) r \right) \tag{34}$$

where $A_s = V_s \exp(\gamma)$, that up to notation ($p = v_r^s / V_s$; $Z^{-1} = \exp(\gamma)$) coincides with (21).

Taking into account (26), one obtains

$$A_s = V_s \frac{1 - \exp\left( (\beta \varepsilon + \alpha) \right)}{1 - \exp\left( (\beta \varepsilon + \alpha)(d+1) \right)} \tag{35}$$

Then, taking into account (28), (34) and (35), the number of elements with a salary $\varepsilon_s$ in the system under consideration is equal to

$$N_s = \sum_{r=0}^{d} r A_s \exp\left( (\beta \varepsilon_s + \alpha) r \right) = V_s f_G(\varepsilon_s) \tag{36}$$

where $f_G(\varepsilon_s)$ is an intermediate Gentile statistics (25).

Using (29), (30), (32) and (36), one obtains

$$N = \sum_s V_s f_G(\varepsilon_s) \tag{37}$$

$$E = -\sum_s \varepsilon_s V_s f_G(\varepsilon_s) \tag{38}$$

$$S = \beta E - \alpha N + \sum_s V_s \ln \frac{1-\exp((\beta\varepsilon+\alpha)(d+1))}{1-\exp((\beta\varepsilon+\alpha))} \tag{39}$$

Let us introduce the distribution $\varphi(\varepsilon)$ of companies by the salary at the level under consideration, where $\varphi(\varepsilon)d\varepsilon$ is the relative number of companies in which the salary on the considered level is in the range $[\varepsilon, \varepsilon+d\varepsilon]$. Then

$$V_s = V\varphi(\varepsilon)d\varepsilon \tag{40}$$

Taking into account (27), one obtains

$$\int_0^\infty \varphi(\varepsilon)d\varepsilon = 1 \tag{41}$$

Then relations (37) - (41) can be written in the form

$$\frac{N}{V} = n \tag{42}$$

$$\frac{E}{N} = u \tag{43}$$

$$S = \beta E - \alpha N + V\omega \tag{44}$$

where

$$n(\alpha,\beta) = \int_0^\infty \varphi(\varepsilon) f_G(\varepsilon)d\varepsilon \tag{45}$$

$$u(\alpha,\beta) = -\frac{\int_0^\infty \varepsilon\varphi(\varepsilon) f_G(\varepsilon)d\varepsilon}{\int_0^\infty \varphi(\varepsilon) f_G(\varepsilon)d\varepsilon} \tag{46}$$

$$\omega(\alpha,\beta) = \int_0^\infty \varphi(\varepsilon) \ln \frac{1-\exp((\beta\varepsilon+\alpha)(d+1))}{1-\exp((\beta\varepsilon+\alpha))} d\varepsilon \tag{47}$$

From relations (42) and (43), taking into account (25) for a given function $\varphi(\varepsilon)$, one can find the dependences:

$$\alpha = \alpha(n,u) \tag{48}$$

$$\beta = \beta(n,u) \tag{49}$$

The relation (44), (47) formally determines the functional dependence $S = S(\alpha,\beta,E,N,V)$, which, taking into account (42), (43), (48) and (49), can be rewritten in the form

$$S = S(E,N,V) \tag{50}$$

The differential of this function

$$dS = \left(\frac{\partial S}{\partial E}\right)_{N,V} dE + \left(\frac{\partial S}{\partial N}\right)_{E,V} dN + \left(\frac{\partial S}{\partial V}\right)_{E,N} dV \qquad (51)$$

We introduce the notation

$$\left(\frac{\partial S}{\partial E}\right)_{N,V} = \frac{1}{T}; \left(\frac{\partial S}{\partial N}\right)_{E,V} = -\frac{\mu}{T}; \left(\frac{\partial S}{\partial V}\right)_{E,N} = \frac{p}{T} \qquad (52)$$

By analogy with thermodynamics, the parameter $T$ defined in this way we will call the temperature of the economic system; the parameter $\mu$ we will call the financial potential of the system, and the parameter $p$ can be called the pressure in the economic system. These parameters characterize the economic system as a whole and should be the same for any subsystem of this system. Obviously, the financial potential in economic systems is an analogue of the chemical potential in physical thermodynamic systems.

Taking into account (52), the relation (51) takes the form

$$TdS = dE - \mu dN + pdV \qquad (53)$$

Expression (53) has the form of the second law of thermodynamics for equilibrium systems. If we introduce

$$\delta Q = TdS \qquad (54)$$

equation (53) takes the form of the first law of thermodynamics for a system with a variable number of elements:

$$\delta Q = dE - \mu dN + pdV \qquad (55)$$

As in the conventional physical thermodynamic system, for the economic system under consideration, all parameters can be separated into extensive ones (varying proportionally to the number of elements in the system) and intensive ones, which depend only on the ratio of extensive parameters or on other intensive parameters. It follows from the definition (42) - (47) that the volume of the system, its energy and entropy are extensive parameters of the system. According to the definition (52), the pressure and the financial potential of the system are intensive parameters of the system.

We rewrite relation (53) in the form

$$dG = -SdT + \mu dN + Vdp \qquad (56)$$

where

$$G = E - TS + pV \qquad (57)$$

is the Gibbs free energy of the economic system.

It follows from (56) that

$$G = G(T, N, p) \qquad (58)$$

and is the extensive parameter.

From here, as usual, it is easy to obtain

$$G = N\mu(T, p) \tag{59}$$

Then expression (57) can be rewritten in the form

$$E - TS - \mu N = -pV \tag{60}$$

The relation (60) is formally similar to the relation (44). For this reason, in thermostatistics it is usually assumed $\beta = 1/T$; $\alpha = \mu/T$. However, it is easy to see that in the general case this is not so. Indeed, according to (52), we have the conditions

$$\frac{\partial}{\partial N}(1/T) = -\frac{\partial}{\partial E}(\mu/T); \frac{\partial}{\partial V}(1/T) = \frac{\partial}{\partial E}(p/T); \frac{\partial}{\partial N}(p/T) = -\frac{\partial}{\partial V}(\mu/T) \tag{61}$$

Obviously, the parameters $\alpha$ and $\beta$, determined by the relations (48) and (49), in the general case do not satisfy relations (61).

Taking into account that entropy is an extensive parameter, we can rewrite relation (50) in the form

$$S = N\psi(n, u) \tag{62}$$

where according to (44)

$$\psi = \omega/n + \beta u - \alpha \tag{63}$$

Then the definitions (52) can be rewritten in the form

$$\frac{1}{T} = \frac{\partial \psi}{\partial u} \tag{64}$$

$$\frac{\mu}{T} = u\frac{\partial \psi}{\partial u} - \frac{\partial n\psi}{\partial n} \tag{65}$$

$$\frac{p}{nT} = -n\frac{\partial \psi}{\partial n} \tag{66}$$

Relations (63) - (66) determine the temperature, pressure and financial potential of the system as a function of the parameters $\alpha$ and $\beta$ or $n$ and $u$.

Taking (48) and (49) into account, equation (66) can be rewritten in the form

$$\frac{pV}{NT} = f(n, T) \tag{67}$$

This equation can be called the thermal equation of state of the economic system.

Similarly, the relation (43), (46) taking into account (25) determines the dependence

$$E = Nu(T, n) \tag{68}$$

Relation (68) can be called the energy equation of state of the economic system.

To calculate the derivatives $\frac{\partial \psi}{\partial u}$ and $\frac{\partial \psi}{\partial n}$ we use the expressions

$$\frac{\partial \psi}{\partial u} = \left(\frac{\partial \psi}{\partial u}\right)_{\alpha,\beta} + \left(\frac{\partial \psi}{\partial \alpha}\right)_{u,\beta} \frac{\partial \alpha}{\partial u} + \left(\frac{\partial \psi}{\partial \beta}\right)_{u,\alpha} \frac{\partial \beta}{\partial u} \tag{69}$$

$$\frac{\partial \psi}{\partial n} = \left(\frac{\partial \psi}{\partial n}\right)_{\alpha,\beta} + \left(\frac{\partial \psi}{\partial \alpha}\right)_{n,\beta} \frac{\partial \alpha}{\partial n} + \left(\frac{\partial \psi}{\partial \beta}\right)_{n,\alpha} \frac{\partial \beta}{\partial n} \tag{70}$$

Taking into account (63), one obtains

$$\frac{\partial \psi}{\partial u} = \beta + \left[\frac{1}{n}\left(\frac{\partial \omega}{\partial \alpha}\right) - 1\right]\frac{\partial \alpha}{\partial u} + \left[\frac{1}{n}\left(\frac{\partial \omega}{\partial \beta}\right) + u\right]\frac{\partial \beta}{\partial u} \tag{71}$$

$$\frac{\partial \psi}{\partial n} = -\frac{\omega}{n^2} + \left[\frac{1}{n}\left(\frac{\partial \omega}{\partial \alpha}\right) - 1\right]\frac{\partial \alpha}{\partial n} + \left[\frac{1}{n}\left(\frac{\partial \omega}{\partial \beta}\right) + u\right]\frac{\partial \beta}{\partial n} \tag{72}$$

To calculate the derivatives $\frac{\partial \alpha}{\partial n}, \frac{\partial \alpha}{\partial u}, \frac{\partial \beta}{\partial n}$ and $\frac{\partial \beta}{\partial u}$ we use the properties of Jacobian $J = \frac{\partial(n,u)}{\partial(\alpha,\beta)}$, assuming that $J \neq 0$. As a result, one obtains

$$\frac{\partial \alpha}{\partial n} = J^{-1}\frac{\partial u}{\partial \beta}; \quad \frac{\partial \alpha}{\partial u} = -J^{-1}\frac{\partial n}{\partial \beta}; \quad \frac{\partial \beta}{\partial n} = -J^{-1}\frac{\partial u}{\partial \alpha}; \quad \frac{\partial \beta}{\partial u} = J^{-1}\frac{\partial n}{\partial \alpha} \tag{73}$$

Then (71) and (72) take the form

$$\frac{\partial \psi}{\partial u} = \beta - J^{-1}\left[\frac{1}{n}\left(\frac{\partial \omega}{\partial \alpha}\right) - 1\right]\frac{\partial n}{\partial \beta} + J^{-1}\left[\frac{1}{n}\left(\frac{\partial \omega}{\partial \beta}\right) + u\right]\frac{\partial n}{\partial \alpha} \tag{74}$$

$$\frac{\partial \psi}{\partial n} = -\frac{\omega}{n^2} + J^{-1}\left[\frac{1}{n}\left(\frac{\partial \omega}{\partial \alpha}\right) - 1\right]\frac{\partial u}{\partial \beta} - J^{-1}\left[\frac{1}{n}\left(\frac{\partial \omega}{\partial \beta}\right) + u\right]\frac{\partial u}{\partial \alpha} \tag{75}$$

Using (47) and taking into account (25), (45) and (46), one obtains

$$\frac{\partial \omega}{\partial \alpha} = \int_0^\infty \frac{\partial \varphi(\varepsilon)}{\partial \alpha} \ln \frac{1-\exp((\beta\varepsilon+\alpha)(d+1))}{1-\exp((\beta\varepsilon+\alpha))} d\varepsilon + n \tag{76}$$

$$\frac{\partial \omega}{\partial \beta} = \int_0^\infty \frac{\partial \varphi(\varepsilon)}{\partial \beta} \ln \frac{1-\exp((\beta\varepsilon+\alpha)(d+1))}{1-\exp((\beta\varepsilon+\alpha))} d\varepsilon - un \tag{77}$$

Here it is assumed that the function $\varphi(\varepsilon)$ in the general case can depend on the parameters $\alpha$ and $\beta$.

The relations (74) and (75), taking (76) and (77) into account, take the form

$$\begin{aligned}\frac{\partial \psi}{\partial u} &= \beta - \frac{1}{nJ}\frac{\partial n}{\partial \beta}\int_0^\infty \frac{\partial \varphi(\varepsilon)}{\partial \alpha} \ln \frac{1-\exp((\beta\varepsilon+\alpha)(d+1))}{1-\exp((\beta\varepsilon+\alpha))} d\varepsilon + \\ &+ \frac{1}{nJ}\frac{\partial n}{\partial \alpha}\int_0^\infty \frac{\partial \varphi(\varepsilon)}{\partial \beta} \ln \frac{1-\exp((\beta\varepsilon+\alpha)(d+1))}{1-\exp((\beta\varepsilon+\alpha))} d\varepsilon\end{aligned} \tag{78}$$

$$\frac{\partial \psi}{\partial n} = -\frac{\omega}{n^2} + \frac{1}{nJ}\frac{\partial u}{\partial \beta}\int_0^\infty \frac{\partial \varphi(\varepsilon)}{\partial \alpha}\ln\frac{1-\exp((\beta\varepsilon+\alpha)(d+1))}{1-\exp((\beta\varepsilon+\alpha))}d\varepsilon -$$
$$-\frac{1}{nJ}\frac{\partial u}{\partial \alpha}\int_0^\infty \frac{\partial \varphi(\varepsilon)}{\partial \beta}\ln\frac{1-\exp((\beta\varepsilon+\alpha)(d+1))}{1-\exp((\beta\varepsilon+\alpha))}d\varepsilon \qquad (79)$$

If the function $\varphi(\varepsilon)$ does not depend on the parameters $\alpha$ and $\beta$, one obtains

$$\frac{\partial \psi}{\partial u} = \beta \qquad (80)$$

$$\frac{\partial \psi}{\partial n} = -\frac{\omega}{n^2} \qquad (81)$$

In this case, taking (64) - (66) into account, we obtain the usual relation between statistics and thermodynamics:

$$\beta = \frac{1}{T} \qquad (82)$$

$$\alpha = \frac{\mu}{T} \qquad (83)$$

$$\frac{p}{T} = \omega \qquad (84)$$

Thus, we see that the usually postulated relation between the parameters $\alpha$ and $\beta$ in the Gibbs distribution and the thermodynamic temperature $T$ and potential $\mu$ is not universal, and it only occurs when the function $\varphi(\varepsilon)$ does not depend on the parameters $\alpha$ and $\beta$.

In the general case, all parameters (42) - (47) depend on the specific form of the function $\varphi(\varepsilon)$.

As an example, we consider the case when

$$\varphi(\varepsilon) = \delta(\varepsilon - \varepsilon_0) \qquad (85)$$

where $\varepsilon_0$ is the constant parameter.

Condition (85) means that the salary at a given level for all companies is the same and equal to $\varepsilon_0$.

In this case

$$n = f_G(\varepsilon_0) \qquad (86)$$

$$\omega = \ln\frac{1-\exp((\beta\varepsilon_0+\alpha)(d+1))}{1-\exp((\beta\varepsilon_0+\alpha))} \qquad (87)$$

where the parameters $\alpha$ and $\beta$ are related to the temperature and financial potential of the system by the relations (82) and (83), while the pressure in the system is determined by the relation (84).

The equation of state of system (67) is given parametrically by the relations (82), (84), (86) and (87). It is convenient to depict it in coordinates $p/T$, $N/(Vd)$. Fig. 5 shows the equations of state of the system under consideration for different levels $d$ under the condition (85).

We see that for small values of the parameter $N/(Vd)$ with sufficient accuracy the parameter $p/T$ depends linearly on $N/V$.

It follows from Fig. 5, that the small values of $\dfrac{N}{V} \ll 1$ correspond to $\lambda \ll -1$, where $\lambda = \beta \varepsilon_0 + \alpha$. In this case $\exp(-\lambda(d+1)) \gg 1$ and $\exp(-\lambda(d+1)) \gg \exp(-\lambda)$, moreover, the second inequality is satisfied the better, the larger $d$. Then, using (25), (84), (86), and (87), we approximately obtain:

$$\frac{N}{V} \approx \frac{1}{\exp(-\lambda)-1}; \quad p \approx T \ln \frac{1}{1-\exp(\lambda)}$$

or

$$p \approx T \ln(1 + N/V) \qquad (88)$$

At $N/V \ll 1$, one obtains

$$pV \approx NT \qquad (89)$$

Thus, when $N/V \ll 1$ the economic system is approximately described by the Mendeleev-Clapeyron equation (ideal gas law). Such an economic system can be called ideal (primitive) by analogy with a gas satisfying the equation of state (89).

The dependence of the financial potential of the system on the ratio $N/V$ is conveniently represented in coordinates $N/(Vd)$, $(\varepsilon_0 + \mu)/T$. Such dependences for different levels $d$ under condition (85) are shown in Fig. 6.

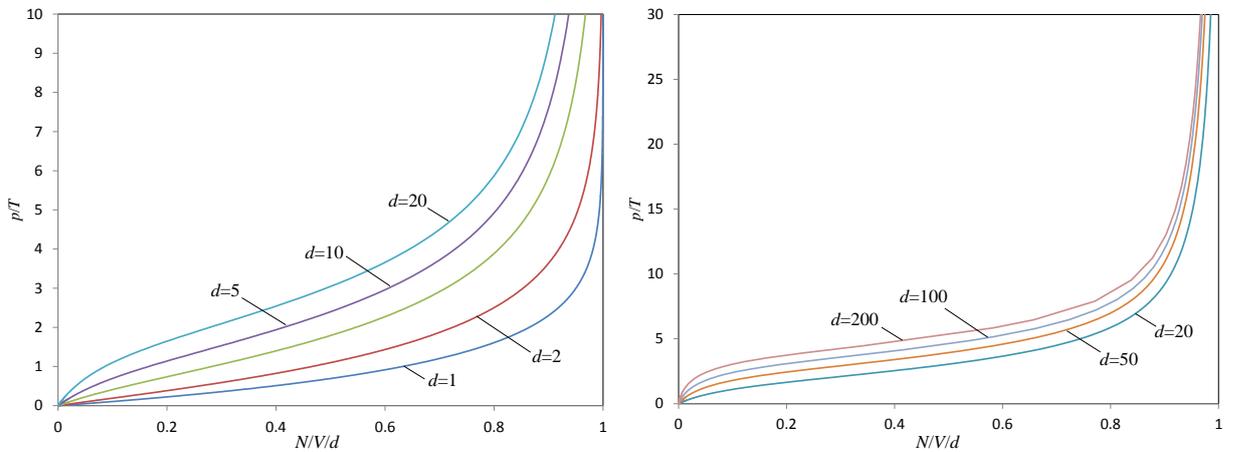

Fig. 5. The thermal equation of state $p = p(T, N/V)$ of the economic system for different levels $d$ under condition (85).

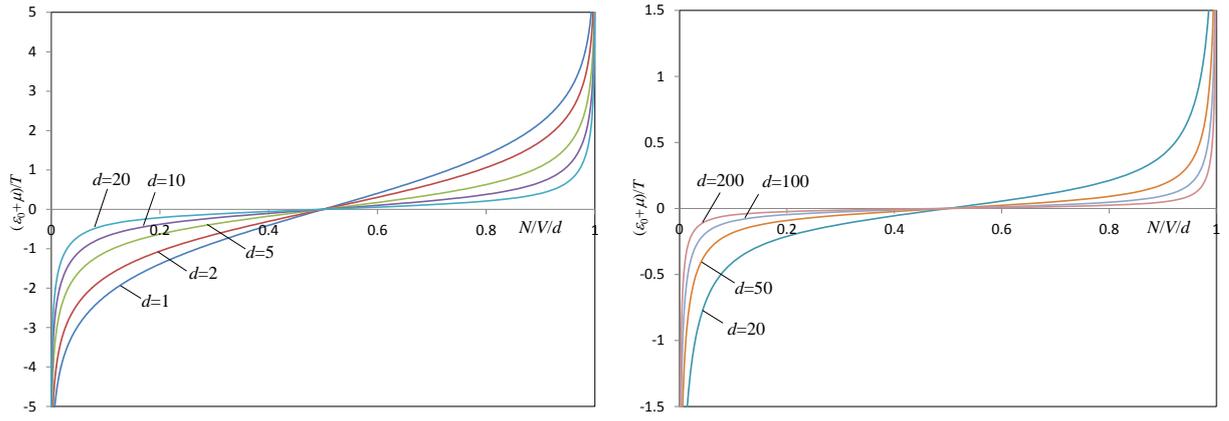

Fig. 6. Dependence of the financial potential of the system on the ratio $N/V$ for different levels $d$ under condition (85).

We see that, for example, for $d = 50000$ at large temperatures the degree of population of the level is small: $N/(Vd) \ll 1$. As the temperature decreases from $(T/p)\ln(d+1) = 1.25$ to $(T/p)\ln(d+1) = 0.6$ the degree of population increases sharply from $N/(Vd) = 0.1$ to $N/(Vd) = 0.9$, and at $(T/p)\ln(d+1) < 0.6$, more than 90% of the positions at this level are occupied. As the parameter $d$ increases, the temperature range in which a sharp change in the population of the level occurs is narrowed. Such an abrupt change in the degree of population of the level can be attributed to the Bose-Einstein condensation in the system under consideration: when the temperature decreases below a certain value, 'condensation' occurs, as a result of which most of the positions of this level are occupied. On the contrary, if the temperature exceeds a certain value, 'evaporation' occurs, as a result of which the majority of the positions of this level turn out to be vacant. Of course, in this case condensation does not occur suddenly when passing through a certain critical temperature, but continuously; however, this change in the system occurs in the rather narrow temperature range, which narrows as the parameter $d$ increases. Nevertheless, it is possible to conditionally introduce a critical temperature characterizing the onset of intense condensation. As such a critical temperature, we can take, for example, the temperature at which the population of the considered level is 50% of the maximum possible, i.e. when $N/(Vd) = 0.5$. Then the critical temperature of condensation is

$$T_c = \frac{p}{\ln(d+1)}.$$

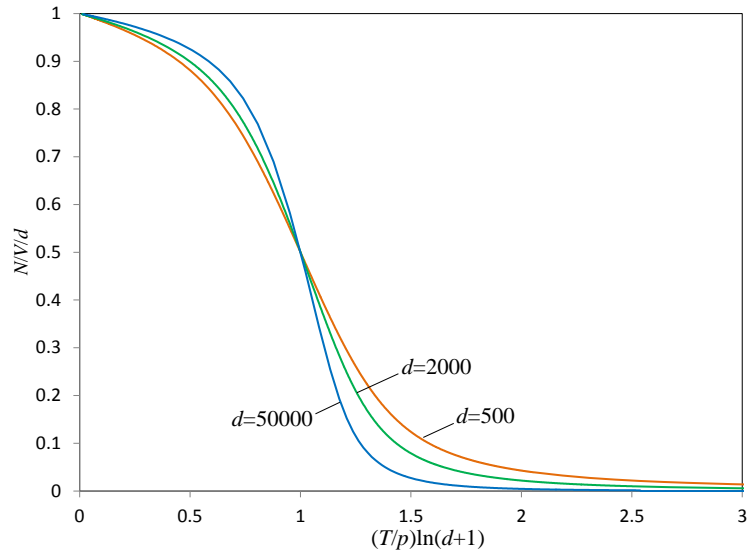

Fig. 7. Dependence of $N/(Vd)$ on $(T/p)\ln(d+1)$ for large values of the parameter $d \gg 1$ under condition (85).

## 6. Concluding remarks

Thus, we see that for social and economic systems that have a hierarchical structure, it is possible to construct thermostatistics, which is based on the intermediate Gentile statistics [10, 11]. The elements of such hierarchical systems, depending on the level, have the properties of either fermions or bosons, or they have an intermediate properties between fermions and bosons. The analysis shows that such concepts as temperature, pressure and financial potential, which is an analogue of the chemical potential in a thermodynamic system, can be introduced for economic systems. At present, the economic meaning of these parameters is unclear, but one can expect that they are related to other well-known indicators of economic systems.

In our opinion, the thermodynamic description of economic systems seems promising; it will allow analyzing the states of economic systems and economic processes (including cyclical ones), using methods well developed for physical thermodynamic systems. From this point of view, the analysis of economic processes is similar to the analysis of the operation of thermal machines. This issue will be considered in the subsequent articles.

In conclusion, we point out two interesting analogies that arose in the course of the analysis.

In [12], the idea of a 'social laser' was put forward: action amplification by stimulated emission of social energy. This idea gets a concrete meaning with reference to the hierarchical system considered in this article, which can be represented as a potential well with discrete levels of energy (salary). If the system is stimulated by external impact, i.e. a number of elements are transferred to a higher energy level (a level with a lower salary), leaving positions with a higher salary (lower energy) vacant, then a population inversion appears in the system. In such a state, the system without an external influence is unstable. After removing the external impact, a rapid spontaneous transition of elements from a higher energy level to a lower one will occur in the system there, which will lead to a drastic change in the system (up to its destruction).

Another interesting analogy relates to salaries in a hierarchical economic system. At each level of the hierarchical system, the salary is different, moreover, the salary changes from one level to another not continuously, but abruptly. In other words, there is a 'discrete spectrum' of salaries in the hierarchical system, and the higher the energy level (the less salary), the less the difference in salaries at neighboring levels. This resembles the discrete energy spectrum of a quantum particle in a potential well (for example, an electron in an atom).

It is well-known that the possible energy levels of a quantum system are the eigenvalues of some linear operator (Hamiltonian). The question arises: is it possible to construct an operator whose eigenvalues are salaries at different levels of the hierarchical economic (social) system?

**Acknowledgements**

Funding was provided by the Tomsk State University competitiveness improvement program.

**References**

[1] E. Haven, A. Khrennikov, T. Robinson, *Quantum Methods in Social Science: A First Course.* World Scientific Publishing Company, 2017.

[2] E. Haven, A. Khrennikov, eds., *Applications of Quantum Mechanical Techniques to Areas Outside of Quantum Mechanics*. Lausanne: Frontiers Media, 2018. doi: 10.3389/978-2-88945-427-3.

[3] B. K. Chakrabarti, A. Chakraborti, S. R.Chakravarty, A. Chatterjee*, Econophysics of income and wealth distributions*. Cambridge University Press, 2013.

[4] A. Dragulescu, V.M. Yakovenko, Statistical mechanics of money, V. Eur. Phys. J. B **17** (2000) 723.

[5] A. Drăgulescu, V.M. Yakovenko, Exponential and power-law probability distributions of wealth and income in the United Kingdom and the United States, Physica A: Statistical Mechanics and its Applications, **299** (2001) 213.

[6]  P. K. Mohanty, Generic features of the wealth distribution in ideal-gas-like markets, Physical Review E **74** (2006) 011117.

[7]  W. P. Cockshott, A. F. Cottrell, I. P. Wright, G. J. Michaelson, V. M. Yakovenko, *Classical econophysics.* Routledge, 2009.

[8]  V. M. Yakovenko, Statistical mechanics approach to the probability distribution of money. *New Approaches to Monetary Theory: Interdisciplinary Perspectives*; H. Ganssmann, Ed, (2012) 104.

[9]  V.M. Yakovenko, Monetary economics from econophysics perspective, The European Physical Journal Special Topics, **225**, (2016) 3313. https://doi.org/10.1140/epjst/e2016-60213-3.

[10] G. Gentile j., itOsservazioni sopra le statistiche intermedie. Il Nuovo Cimento (1924-1942), **17** (1940), 493.

[11] G. Gentile, Le statistiche intermedie e le proprieta dell'elio liquid, Il Nuovo Cimento (1924-1942), **19** (1942) 109.

[12] A. Khrennikov, 'Social Laser': action amplification by stimulated emission of social energy. Phil. Trans. R. Soc. A, **374** (2016) 20150094.